# Comparison between Idealized and Realized Accessibility Measures in GIScience

Yuyan Che[1], Trisalyn A. Nelson[1], Somayeh Dodge[1], Peter Kedron[1]

yuyanche@ucsb.edu, trisalyn@ucsb.edu, sdodge@ucsb.edu, peterkedron@ucsb.edu

1. Department of Geography, University of California Santa Barbara, Santa Barbara, CA, USA

0009-0000-5835-5515, 0000-0003-2537-6971, 0000-0003-0335-3576, 0000-0002-1093-3416

## Abstract

Measures of access, defined as the ease with which people can reach opportunities or services, are often based on proximity. Proximity measures of access are often unrealistic or idealized, ignoring many of the real barriers to access including social and economic barriers. There is a need to develop GIS measures of access that incorporate all aspects of access, which we term realized access. Our goal is to develop a conceptual framework of realized accessibility in active transportation, and compare idealized and realized measures of access to bicycling. We apply the framework to measure realized accessibility in a case study focusing on bicycling access in Santa Barbara County, and compare idealized and realized measures of access for four cities in Santa Barbara County in California (Santa Maria, Lompoc, Santa Barbara, and Goleta). Differences in measures from idealized access to realized access are greatly increased in cities with lower median household income and a higher proportion of people identifying as Hispanic. Studies aiming to understand equity and access can benefit from more nuanced and realized access measures, as idealized access measures may overestimate accessibility in underserved communities. In GIScience, especially as new data on mobility become more widely available, nuanced measures of accessibility should be the standards in analysis.

## 1. Introduction

There have been a growing number of studies (e.g., Geertman and Ritsema Van Eck 1995, Liu and Zhu 2004, Pot et al. 2021, Cracu et al. 2024) that leverage measures of accessibility to characterize who has access to facilities and services and how this varies over space, time, and populations. Accessibility studies have been common in GIScience as researchers measure access to food (Bradley and Vitous 2019), health care (Wang 2020), and destinations (Cheng and Bertolini 2013). In GIScience, measures of access are generally defined by proximity, and consider nearby facilities or services accessible (Wachs and Kumagai 1973). However, proximity measures of access can be overly simplistic and are *idealized*, assuming nearness is the only factor of accessibility (Pearce et al. 2006, La Rosa 2014, Banerjee et al. 2020). These measures of *idealized access* overlook social and economic factors that impact accessibility.

More inclusive, or *realized access* measures should be created and extend beyond mere proximity to encompass all the opportunities and constraints that determine true access (Zheng et al. 2024), which include natural environment, built environment, socio-economic factors, and psychological factors (Heinen et al. 2010, Eckert and Shetty 2011). In the example of food access, measures of *realized access* to food require a reachable distance to food outlets, safe travel from origin to destination, and sufficient funds to get food. While *idealized access* measures focus on the population with access to the facilities or services, *realized access* measures consider those who actually use the facilities or services.

The emergence of new data on mobility enriched with socio-demographic and contextual information creates an opportunity to address the long-standing lack of information needed to develop measures of actual or *realized* accessibility. Modern mobility datasets, such as cell phone GPS based mobility data (e.g., Replica (Chow and Ren 2023)), fitness and movement application data (e.g., Strava (Ferster et al. 2021)), and points of interest (POI) data (e.g., SafeGraph (Juhasz and Hochmair 2020)), enable new forms of measurement. By leveraging spatial patterns that can be captured with these data, we can better represent the processes and contextual factors of accessibility and barriers to accessibility.

One area where novel mobility datasets can be used to improve the accuracy measure accessibility by evaluating differences between measures of *idealized* and *realized access* is active transportation (AT), defined as human-powered, non-motorized mode of transportation (i.e., bicycling and walking) (Mueller et al. 2015). Active transportation accessibility, especially bike related measures, are often *idealized*. For example, it is common to measure accessibility of bicycling as a persons' proximity to bicycle infrastructure, whereby all people residing within a certain distance (e.g., 500 m) of a bike lane are considered to have access (Hochmair 2015, Hosford and Winters 2018, Weliwitiya et al. 2019). Cities may even measure progress in transportation based on the percentage of people living proximal to bike lanes. However, access to bicycling is more than proximity to bike lanes and can be related to access to destinations, and other individual-level factors such as gender, age, and race/ethnicity (Barajas 2017, Frater and Kingham 2018, Grudgings et al. 2021). As the public interest in biking grows, measures of access for this sustainable and healthy commuting mode (Gilderbloom et al. 2016, Wild and Woodward 2019) need to be redefined.

In this research, we lay the foundation for a conceptual framework of *realized* accessibility in active transportation, and compare *idealized* and *realized access* measures to bicycling. To approach to the true access, we apply the framework to measure *realized* accessibility in a case study focusing on bicycling access in Santa Barbara County, and compare measures of *idealized* and *realized access* for four cities in Santa Barbara County in California (Santa Maria, Lompoc, Santa Barbara, and Goleta). While the methods developed are specific to the application of bicycling, the broader concept and framework of *realized* accessibility is widely applicable across GIScience measures of access.

## 2. Towards A Conceptual Framework of Accessibility in Active Transportation

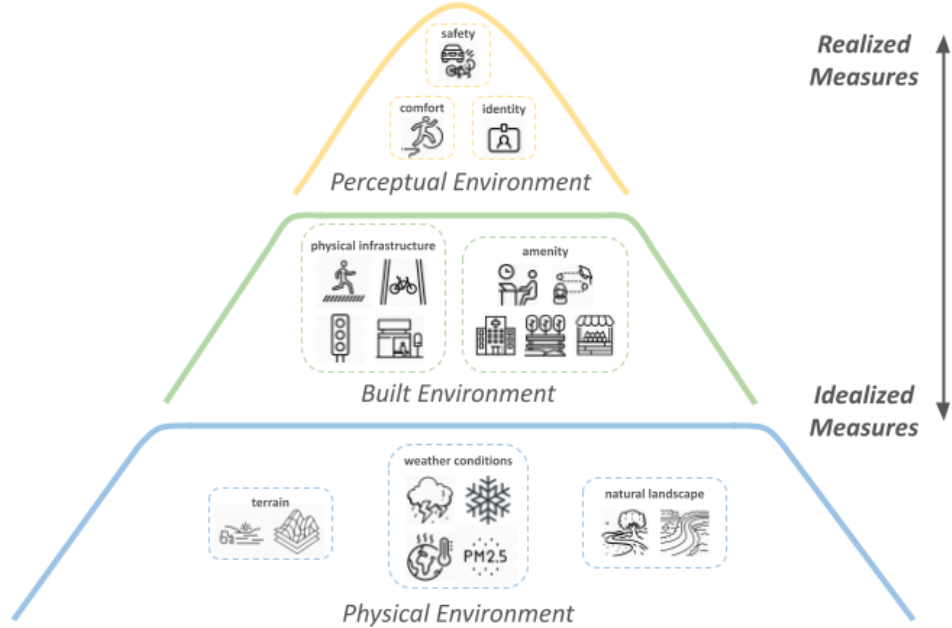

**Fig. 1.** Active transportation *realized* accessibility framework foundation

Conventional or i*dealized access* measures emphasize proximity to physical and built environment, while overlooking social, economic, or cultural factors that impact individuals' accessibility. *Idealized access* measures establish buffers around transportation facilities (such as bus stations and bike lanes) and quantify the proportion of the population that falls within the service area (O'Neill et al. 1992, Gutiérrez and García-Palomares 2008, El-Geneidy et al. 2014, Hochmair 2015). Other measures are evaluated by assessing network-based or Euclidean distances that individuals need to travel from population centers to reach various facilities (Chen et al. 2007, Delamater et al. 2012, Buczkowska et al. 2019).

The modern time geography framework for individualized access (Dodge and Nelson 2023) provides a foundation for a conceptual framework of *realized* accessibility. The modern framework for time geography emphasizes that people experience access at an individual level and identifies the importance of including demographic factors (e.g., age, gender, ethnicity, household income level, and education attainment) (Piatkowski and Marshall 2015, Zhao et al. 2018, Buehler et al. 2020) in measures of access, which we consider *realized access* measures. In this paper, we extend the Dodge and Nelson's (2023) framework for active transportation (AT), specifically bicycling and walking, and name it the AT *realized* accessibility framework.

Here we lay the foundation for an AT *realized* accessibility framework based on conceptual *idealized* factors that influence an individual's access to bicycling and walking. Factors are categorized into three groups: physical environment (natural constraints and benefits), built environment (human-made infrastructure), and perceptual environment (psychological and socioeconomic aspects). In future research, GIS data structure, tools, and standards may leverage our work to deploy an integrated foundation for measuring AT *realized* accessibility.

## 2.1. Physical environment

In the active transportation field, the physical environment can create natural constraints that can influence the accessibility of walking and biking (Wendel-Vos et al. 2004). Natural constraints may result from terrain (e.g., hills, slopes), weather conditions (e.g., rain, snow, temperature, air quality), and natural landscape (e.g., water features, greenery). The physical environment serves as the foundational layer in the AT *realized* accessibility framework.

Many of the previous studies have revealed the relationship between natural environments and access to walking and bicycling. Hills are known to be negatively associated with walking and biking for commuting (Bopp et al. 2013), and places with more days of precipitation or freezing temperatures are associated with lower usage of utilitarian bicycling (Winters et al. 2007, Zhao et al. 2019). In some cases, physical environmental conditions influence not just who bikes but how often rides are taken, with moderate temperatures being linked to higher travel durations and longer distances for bikeshare trips (Heaney et al. 2019) and greater use of urban trails when conditions are mild (Lanza et al. 2022). The presence of greenery (Schipperijn et al. 2013, Stefansdottir et al. 2019) and pleasant natural spaces (Jansen et al. 2017) are also associated with more walking and biking.

## 2.2. Built environment

The built environment in the AT *realized* accessibility framework represents the human-made or constructed surroundings in which people live, work, and move (Vagegas 2003). We include physical infrastructure (e.g. sidewalks, bike lanes, traffic calming measures, public transit integration), and amenities (e.g. workplaces, schools, hospitals, parks, and markets) that shape urban and rural landscapes and influence human behavior and activities.

The built environment plays a critical role in determining the feasibility, safety, and attractiveness of walking, bicycling, and other non-vehicular modes of travel.

Numerous studies have examined how built environment factors, especially physical infrastructure, affect access to walking and biking. Higher density of sidewalk and bike lanes are associated with more walking and biking (Le et al. 2018, Eldeeb et al. 2021). Other bicycle network characteristics, including network length, network centrality, coverage, and topography, are also correlated with bicycle ridership (Beck et al. 2024). The distribution of amenities is associated with walking and biking for commuting. Areas with a high concentration of jobs are associated with higher walking and biking volumes (Le et al. 2018), and college towns exhibit higher rates of walking to work (McKenzie 2014). Distances between home and school are the main barriers to walking and biking for children (Stewart 2011), while parks, shops, and social networking sites can encourage children to participate in more walking and biking (Larouche et al. 2013).

## 2.3. Perceptual environment

The perceptual environment in active transportation encompasses individuals' perception and experiences (Heesch and Han 2007, Jensen et al. 2017, Hughey et al. 2022). Individuals' perception of AT is influenced by three categories of factors, including safety (Márquez et al. 2021), comfort (Furth et al. 2023), and identity (Wild et al. 2021). Safety estimates the likelihood of exposure to potential harm while walking or biking, particularly concerning crashes (Evans 2003). Comfort refers to the ease and convenience experienced by individuals engaging in walking and biking, opposed to stress which is commonly used as an evaluation approach of suitability (Blanc and Figliozzi 2016). Identity refers to individuals' social and economic status, such as sex, age, race/ethnicity, educational attainment, working status, and income level (Buehler et al. 2020).

Research has been conducted on perceptual factors and their interactions with each other. Most studies on the influence of perceptual factors on active transportation emphasize perceived threats to safety (Aziz et al. 2018, Mueller and Trujillo 2019). In contrast to observed safety measured through the incidence or rates of deaths and injuries (Branion-Calles et al. 2017, Sener et al. 2021), perceived safety is commonly measured through individuals' self-reported responses gathered via questionnaires or surveys (Ross et al. 2017, Lee 2023). Perception of comfort is another key indicator of choosing active modes of transport (Campos Ferreira et al. 2022) and the ease of bicycling influences the bicycling likelihood (Winters et al. 2011). Perceived safety is also known to vary by individuals' gender and time of day. Females perceive higher levels of crime than males while walking and biking (Heinen 2023), and so they may limit their mobility spatially and temporally at the micro level (Lino and Kanashiro 2024), especially in darker and unknown places.

**2.4. Interplay between three environments**

While we conceptualize three separate environments in our AT *realized* accessibility framework, it is the interplay between these environments that leads to *realized* accessibility. Hills, rain, absence of infrastructure, and lack of bicyclists' identity are factors that could conflate and create many barriers to access. Individuals living in urban and rural areas face different challenges in accessing AT infrastructure and destinations (Müller et al. 2024), with the lack of public transportation in rural areas posing significant barriers to bicycling (Hansen et al. 2015). Perceptual safety and comfort are typically associated with natural and built environments (Ewing and Handy 2009, Arellana et al. 2020) and can vary over time. For example, weather events are identified as factors influencing perceptual comfort (Campos Ferreira et al. 2022), and transportation infrastructure impacts perceptual safety (Reynolds et al. 2009). These three environments work together to form this conceptual framework for assessing individuals' accessibility of walking and biking.

## 3. Case study: Bicycle Accessibility in Santa Barbara County

To demonstrate the utility of the proposed framework, we conduct a case study encompassing four cities within Santa Barbara County. Our methodology integrates 2022 American Community Survey (ACS), OpenStreetMap bike infrastructure data, and Strava origins and destinations (O/D) data. To achieve our goal, we establish the following objectives:

1. First, we measure *idealized access* by calculating the population within 500 meters of a bike lane. The measure of *idealized access* reflects the traditional approach to assessing access as proximity to bicycling infrastructure (O'Neill et al. 1992, Gutiérrez and García-Palomares 2008, El-Geneidy et al. 2014, Hochmair 2015).

2. Second, we measure *realized access* by quantifying commuting bike trips from origins weighted by population. The *realized access* measure provides a more realistic approach of actual usage and access to bicycle infrastructure.

3. Third, we quantify the differences between measures of *idealized* and *realized access*. The results assess the differences between measures of *idealized* and *realized access* in bicycling infrastructure across the study area, and then evaluate these differences relative to census demographics on race and income.

## 3.1. Study area

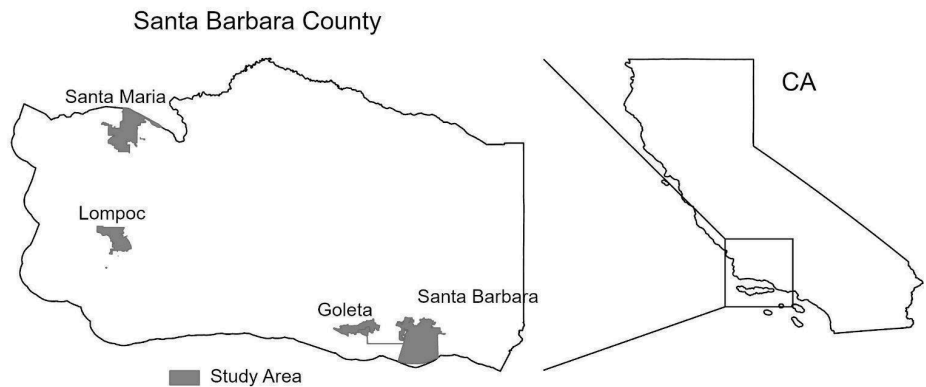

**Fig. 2.** Location of study area, Santa Barbara County

The case study includes four cities located within Santa Barbara County, California: Santa Maria, Lompoc, Goleta, and Santa Barbara (Fig. 2). Santa Barbara County has long supported bicycling and has a rich history of bicycle planning. In 1895, Santa Barbara was the first city in the U.S. to implement a multi-modal application of the new bicycle design by installing bike racks (City of Santa Barbara 1998). In 1974, a landmark bikeway master plan was proposed to increase bike usage in the city (City of Santa Barbara 1998). In 2022, the Active Transportation Program in Santa Barbara County received an $80 million investment from the California Transportation Commission (Gillies 2022).

These four cities in Santa Barbara County exhibit varying levels of bicycling accessibility. Santa Barbara's downtown area ranks among the top 20 cities in the United States with the highest share of bicyclists, boasting a bicycling volume of 4.0% (The League of American Bicyclists 2017). The city offers extensive biking resources, including a comprehensive bike network map, a bike facilities map, a bike share program, and various educational programs aimed to promote bicycling and ensure ride safety. In contrast, Goleta is a more suburban city, adjacent to Santa Barbara, with limited bicycling resources. Santa Maria, with a biking volume of 0.5% in commuting (The League of American Bicyclists 2017), and Lompoc, with an even lower biking volume, are both underserved in bicycle usage.

We also consider two prominent socio-economic factors – median household income and Hispanic population percentage – in order to understand the different spatial patterns in *idealized* and *realized measures* of access in the four cities (Table 1). As a reference, the national median household income is $74,580, and the national Hispanic population accounts for 19.1% of the total population in 2022 (U.S. Census Bureau). Santa Barbara and Goleta have higher median household income than the national average, while Santa Maria and Lompoc have lower median household income than the national average (as summarized in Table 1). All four cities have higher percentages of Hispanic population than the national average, however, Santa Maria and Lompoc have extremely high Hispanic population percentages.

Table 1 Study area city characteristics.

| City | Population | Bike mode share | Median household income | Hispanic population percentage |
|---|---|---|---|---|
| Santa Maria | 114,844 | 0.5% | $67,634 | 76.70% |
| Lompoc | 41,834 | unknown | $57,071 | 60.40% |
| Santa Barbara | 106,887 | 4.0% | $81,618 | 36.70% |
| Goleta | 53,662 | unknown | $98,035 | 33.60% |

*Data source: U.S. Census Bureau, The League of American Bicyclists.

**3.2. Data**

3.2.1. Strava origins and destinations

The crowdsourced 2022 origins and destinations (O/D) data from Strava Metro in shapefile and Excel formats serve as the main dataset for this research focused on measuring *realized access*. The shapefiles delineate the study area into hexagonal grids, with each hexagon covering 0.74 square kilometers (0.28 square miles). The data set represents the origin and destination locations of the Strava biking trips used in this study. Two corresponding Excel files detail trip counts by origins or destinations, specifying the number of trips departing from or arriving at each hexagon. We analyze Strava 2022 the whole year origins and destinations (O/D) data, and both weekdays and weekends data are included. Strava, a physical activity mobile application, tracks users' bicycling records. Although often criticized as biased, Strava data have been validated as representative of overall bicycle volumes in urban areas through comparison with manual bicycling counts (Jestico et al. 2016). The official bicycling data from the government have a limited number of observations, and Strava data address this gap in official government data on bicycle commuting.

Commuting trips refer to travel between an individual's place of residence and workplace (Heinen et al. 2010), identified as commuting by Strava users based on trip purposes. The descriptive statistics for Strava O/D trips (Table 2) are not open to the public due to Strava users' privacy. Among these four cities, both Santa Maria and Lompoc have low numbers of commuting trips and low percentages of Strava users. The city of Santa Barbara has the highest number of commuting trips and the highest percentage of Strava users. The suburban city of Goleta has a medium number of commuting trips and a medium percentage of Strava users.

Table 2 Descriptive statistics for Strava O/D trips.

| City | Strava users | Strava users percentage | Total trips | Commute trips |
|---|---|---|---|---|
| Santa Maria | Lowest** | Lowest | Low | Low |
| Lompoc | Lowest** | Medium low | Lowest | Lowest |
| Santa Barbara | Highest | Highest | Highest | Highest |
| Goleta | Medium low | Medium | Medium | Medium |

*Data source: Strava Metro.

** These values were very similar with only 60 more riders in Santa Maria.

3.2.2. OpenStreetMap and Can-BICS

We use OpenStreetMap (OSM) data and define bike infrastructure following the Canadian Bikeway Comfort and Safety Classification System (Can-BICS) (Winters et al. 2020). We are only interested in bike infrastructure presence and include three categories of Can-BICS classified bike lanes (high, medium, and low comfort bikeways) as conforming bike lanes in our infrastructure data layer. Comfort levels are classified based on stress and safety considerations for bicyclists. The transportation network of Santa Barbara County mainland contains 39,733 polylines with each polyline representing a road, including paths and trails, freeways, and local roads. There are 1,535 bikeways defined as conforming bike lanes in Santa Barbara County. These bike lanes are integrated with 2022 ACS census block data to assess *idealized* accessibility.

### 3.3. Methods

All analyses were primarily conducted using ArcGIS Pro, supplemented by the Python and R programming language.

3.3.1. *Idealized access*

We use the Buffer Analysis tool in ArcGIS Pro to create a 500-meter buffer zone along each bike lane classified through Can-BICS, representing the *idealized* service area for bike infrastructure. Then, we apply the Apportion Polygon tool to allocate population data from the 2022 ACS census blocks into Strava O/D hexagons and buffer zones separately (Fig. 3). For each hexagon, we calculate the proportion of population by dividing the population within the buffer by the total population of the hexagon. This proportional population within the *idealized* service area estimates the percentage of people in proximity to bike infrastructure, referred to as the measure of *idealized access*.

$$idealized\ access\ measure = \frac{population\ within\ 500-meter\ buffer\ zones\ in\ each\ hexagon}{population\ in\ each\ hexagon}$$

(1)

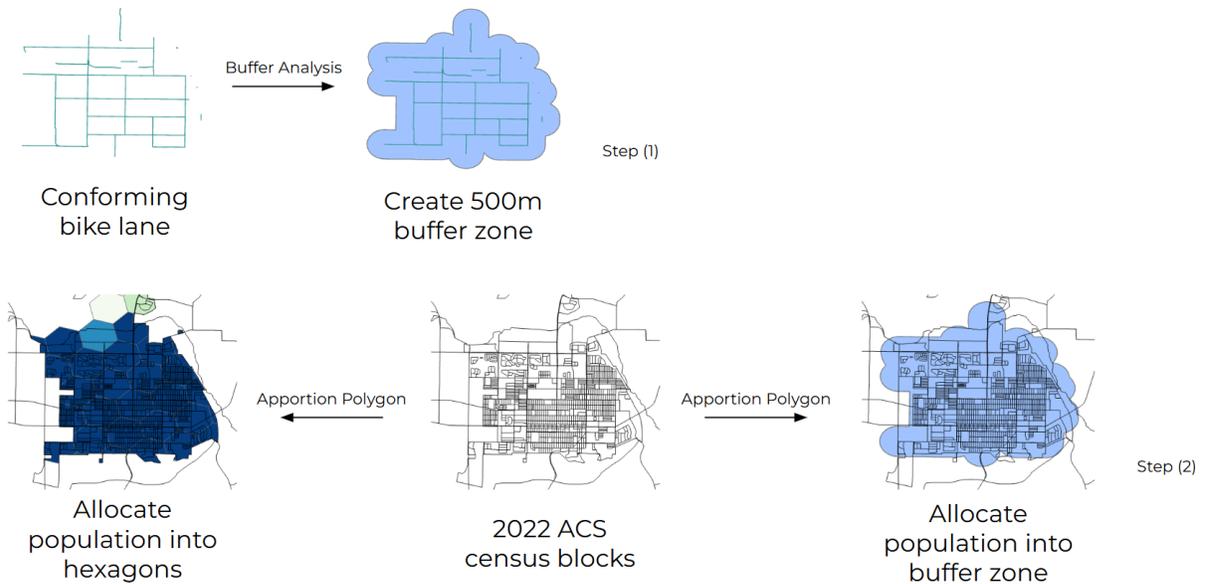

**Fig. 3.** *Idealized access* measures

*3.3.2. Realized access*

In the Strava O/D data, we choose the variable *commuting trips from origins* to measure *realized access*. *Commuting trips from origins* capture the locations of where individuals begin their trips, providing insight into who is using bicycles for commuting. Commuting trips originating from this location reflect the presence of feasible terrain and weather conditions, existing bicycle infrastructure, and individuals' perceptions of safety and comfort. This variable reflects individual commuting behavior, incorporating the underlying factors from the built and physical environments in the framework. We weight *commuting trips from origins* by the population allocated to each hexagon to enable a comparison between *idealized access* and *realized access*. These weighted commuting trips represent the volume of individuals using bike infrastructure, referred to as the measure of *realized access*.

$$realized\ access\ measure = \frac{commuting\ trips\ from\ origins\ in\ each\ hexagon}{population\ in\ each\ hexagon}$$

(2)

3.3.3. Comparative analysis

From Strava O/D data, we identify hexagons with either a non-zero proportional population or weighted commuting trips as valid hexagons. The number of hexagons in the study area cities is as follows: 90 in Santa Barbara, 47 in Goleta, 69 in Santa Maria, and 35 in Lompoc. As stated above, we use the proportional population to represent *idealized access*, and weighted commuting trips to represent *realized access*. Since *idealized* and *realized access* have different units of measurement, they cannot be directly compared. Therefore, we assign ordinal ranks to both *idealized* and *realized access* for each hexagon in ascending order to represent their relative relationships. Ordinal ranking places each value in a specific position within a sequence, with identical values sharing the same rank. Hexagons with the highest number of commuting trips are assigned the largest ordinal rank value, and those with the lowest number of commuting trips are assigned the smallest ordinal rank value. To reflect the transition in measures from *idealized* to *realized* accessibility, we calculate the difference which is reflected by subtracting the rankings of *idealized access* from the rankings of *realized access*.

$$difference\ in\ rank = ranking\ of\ realized\ access - ranking\ of\ idealized\ access$$

(3)

Lange differences in rank values indicate that the hexagon is highly sensitive to how accessibility is measured. If the difference is positive, the ranking of the *realized access* is higher than the ranking of the *idealized access*, indicating that there is a difference between the measures and that accessibility has increased in the *realized access*. Conversely, if the difference is negative, the ranking of the *realized access* is lower than the ranking of the *idealized access*, indicating that there is a difference between the measures and that accessibility has decreased in the *realized access*. If the change equals 0, it indicates that there is no difference in the *idealized access* and *realized access*.

**3.4. Results**

3.4.1. *Idealized access*

The *idealized access* measure is mapped by city and shown in Figure 4, indicating the proportional population within 500 meters along each bike lane. All four cities show a large portion of the population within *idealized* service areas, with most hexagons having over 80% of population in proximity to bike infrastructure. However, there is a notable lack of bike infrastructure in southern Santa Barbara and northern Lompoc.

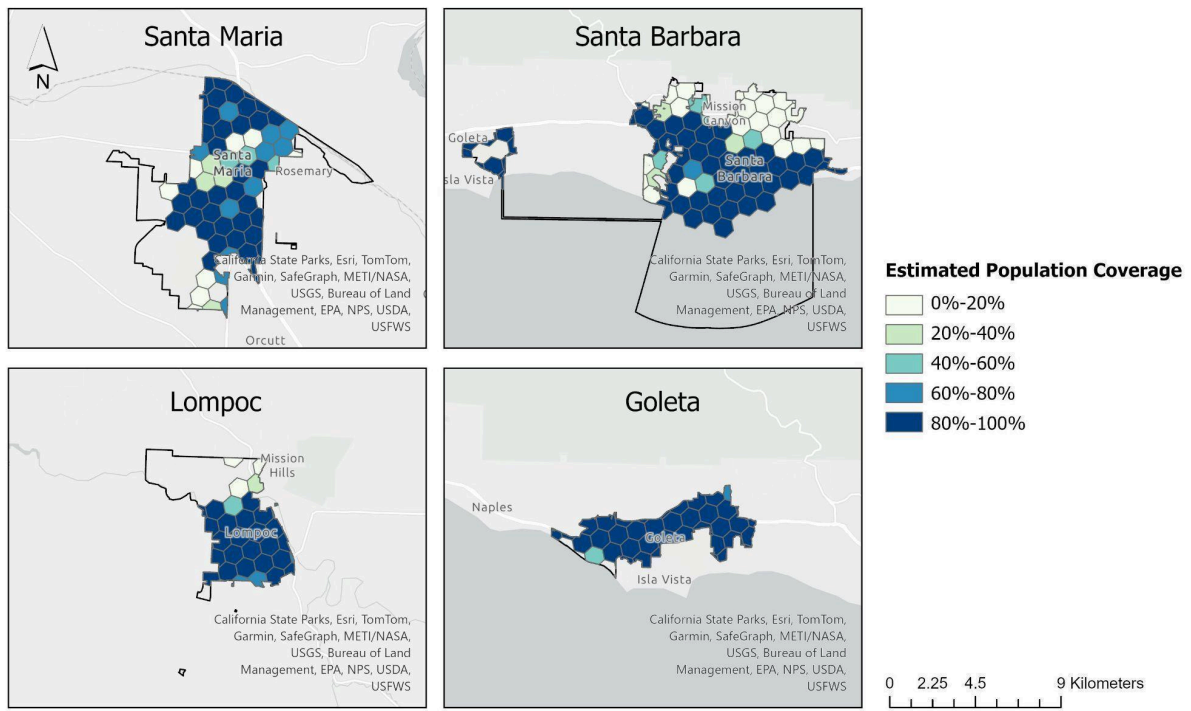

**Fig. 4.** *Idealized access* measure, indicating proportion of population within 500 m buffer of biking infrastructure

3.4.2. *Realized access*

The *realized access* measure is mapped by city and shown in Figure 5, indicating Strava O/D *commuting trips from origins* weighted by population. More than 60% of hexagons in Santa Barbara and Goleta have medium to high weighted commuting trips, which align with the high proportional population in *idealized access*. However, most hexagons in Santa Maria and Lompoc have low weighted commuting trips, which contrast with *idealized access*.

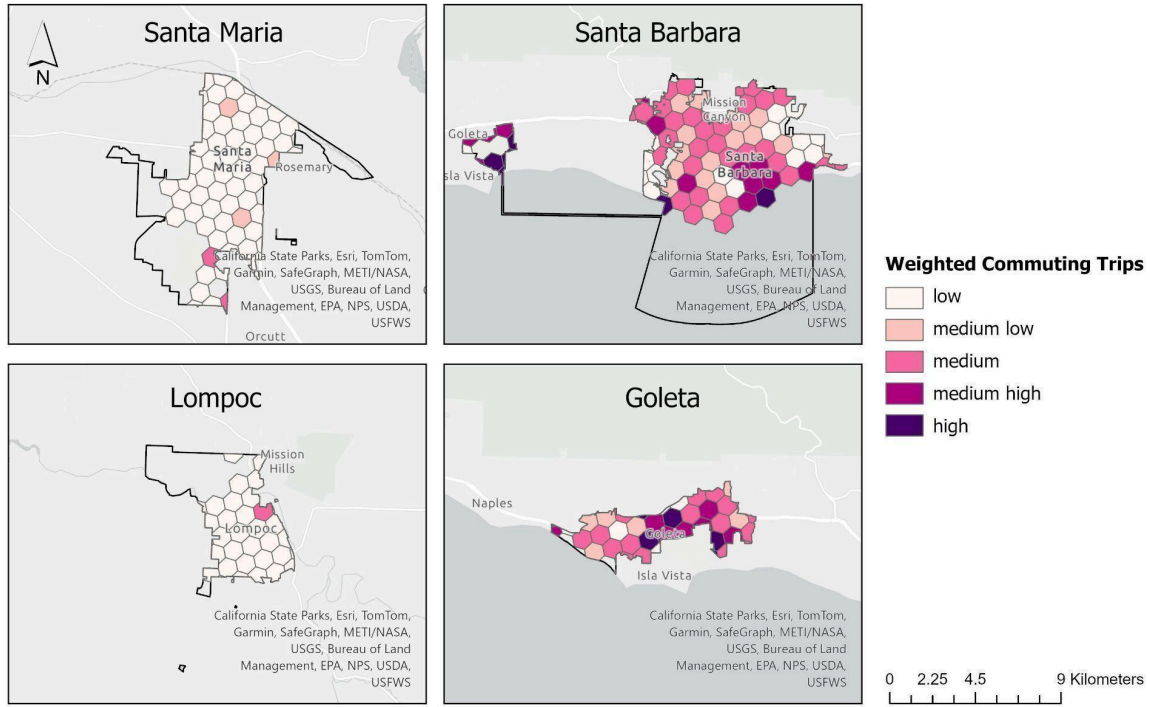

**Fig. 5.** *Realized access* measure, indicating Strava O/D *commuting trips from origins* weighted by population

3.4.3. Comparison of accessibility

The difference in rank between *idealized* and *realized* accessibility for the four cities is mapped by city and shown in Figure 5. Bicycling access in Santa Barbara and Goleta is higher when calculated using *realized access* measures. Most of Santa Barbara showed increased accessibility, with the northern area exhibiting highly increased access (Fig. 6&7). Over 60% of Goleta showed increased access. Only a few hexagons showed no difference between the ranked values of *idealized access* and *realized access* in these two cities. In contrast, bicycling access in Santa Maria and Lompoc is lower when calculated using *realized access* measures. More than 70% of Santa Maria and almost the entire area of Lompoc showed decreased accessibility, with a few hexagons showing no change or increased access. Considering socio-economic factors, increased access is positively correlated with median household income and negatively correlated with the percentage of the Hispanic population (Fig. 8).

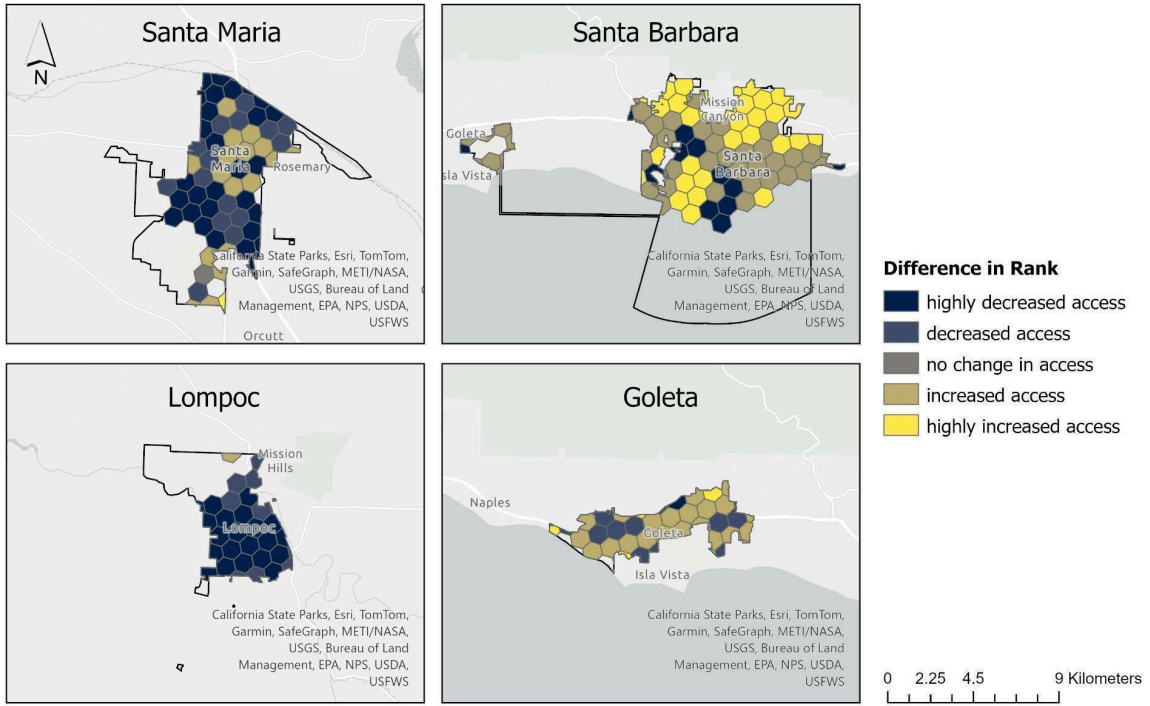

**Fig. 6.** Difference in rank between *idealized* and *realized* accessibility

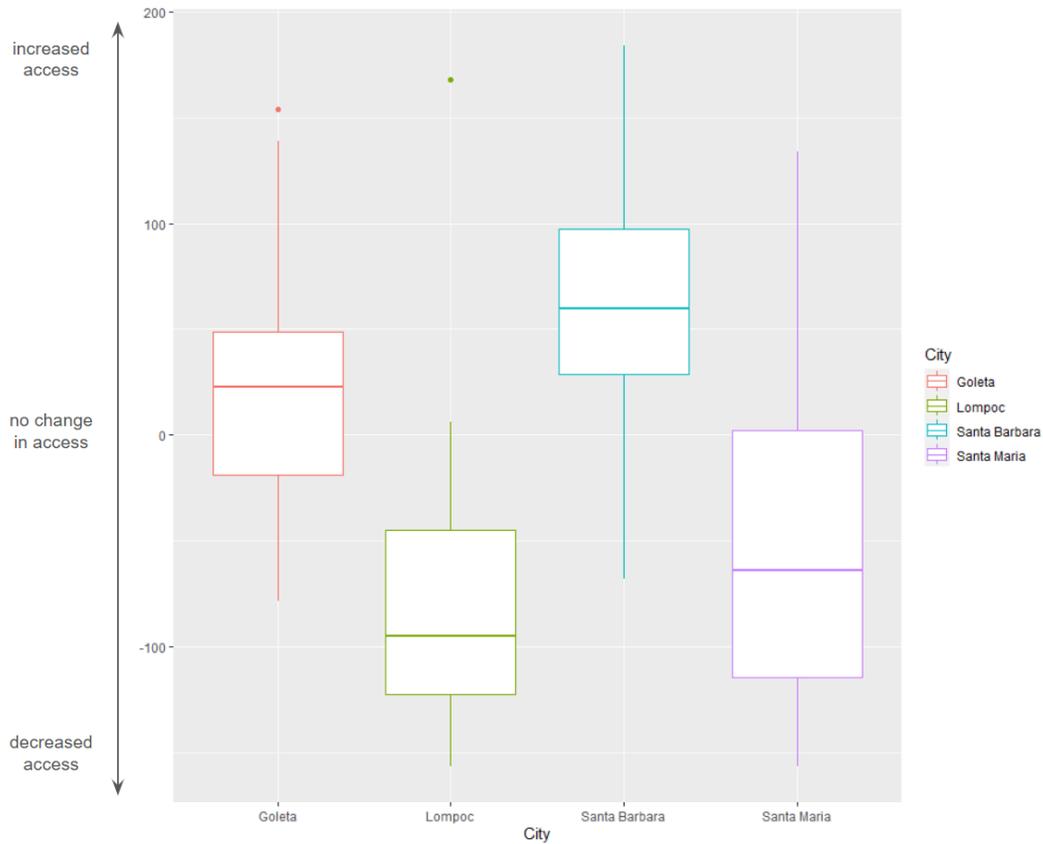

**Fig. 7.** Difference in rank boxplot by cities

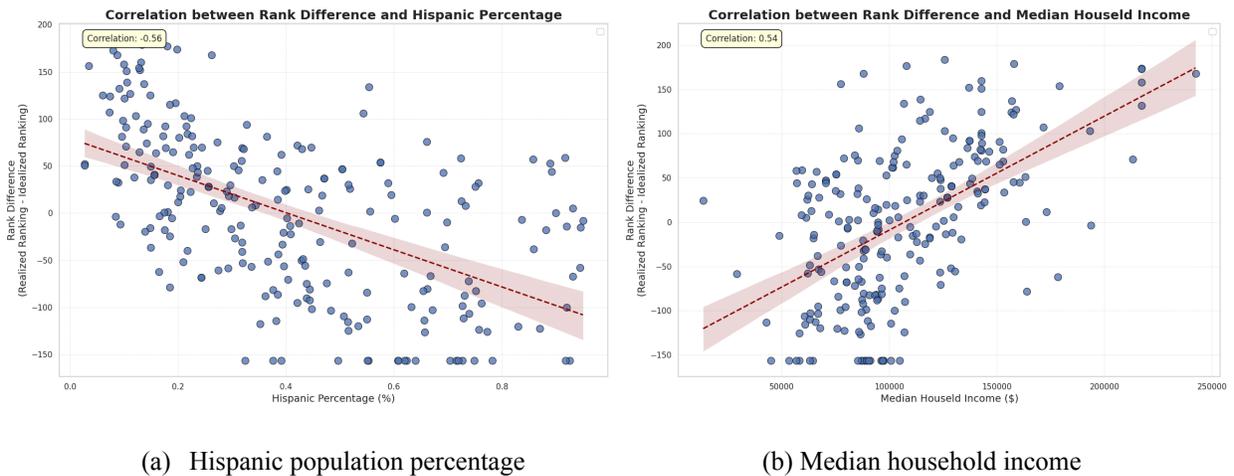

(a)  Hispanic population percentage          (b) Median household income

**Fig. 8.** Correlation between rank difference and socio-economic factors

## 4. Discussion

In GIScience, measures of *idealized access* are based on proximity, which can be simplistic and neglect social and economic factors. Measures of *realized access*, on the other hand, broaden this concept to encompass all the opportunities and constraints related to access, including individuals' socio-economic and psychological factors. The foundation of the AT *realized* accessibility framework categories factors that influence individualized access to bicycling and walking into three environments: physical, built, and perceptual environments. The traditional *(idealized)* approach focuses mainly on physical infrastructure (Vale et al. 2016), whereas the innovative (*realized*) approach incorporates multiple factors from all three environments.

In the case of bicycling, *idealized access* is measured by the population within a 500-meter service area along bike infrastructure which reflects bicycle infrastructure from the built environment. *Realized access* is measured by Strava O/D commuting trips, which emphasizes socio-economic factors from the perceptual environment. Strava commuting trips are influenced by factors across three interconnected environments in the framework. First, terrain and weather conditions must meet the requirements necessary for bicycling, ensuring that the physical environment supports the trip. Second, the presence of accessible bike lanes provides critical routes for cyclists, while the availability of amenities establishes viable origins and destinations. Third, bicyclists are more likely to undertake trips when they perceive the environment as safe and comfortable and free from discrimination based on their identity (Firth et al. 2021, Steinmann et al. 2021, Gupta 2024), highlighting the importance of subjective factors in commuting decisions. The difference between these two access measures reveals that *idealized access* measures can merely reflect the population physically near the bike infrastructure. We propose *realized access* measures to show who actually uses the facilities and further explore equity in accessibility.

One limitation of this study is Strava data are often criticized for its overrepresentation of fitness activities, making it a biased dataset. Unlike existing bicycle accessibility studies that use traditional ridership data (Kabra et al. 2020, Rodriguez-Valencia et al. 2019, Conrow 2021), we use Strava O/D data with commuting trips from origins and destinations. As a novel mobility dataset, Strava O/D data offer a substantial number of observations, surpassing the bicycle volumes provided by the government. In this research, Strava O/D data provide three different types of ridership, including total trips, commuting trips and leisure trips (Table 2). We use commuting trips to reflect overall ridership in *realized access*. Commute ridership samples from Strava have been shown to be a strong predictor of overall bicycle ridership, as the two measures are correlated (Fischer et al. 2020). Using Strava commute data can mitigate the bias of Strava data and more accurately represent overall ridership patterns of people of all ages and abilities. Another limitation of Strava O/D data is the limited information available on commuting trips. While we use the number of commuting trips from origins or destinations, the dataset does not include details such as trip length, duration, or timestamps. Future work could address these gaps by incorporating data on the comfort and safety levels of bike lanes, as well as additional identity information such as age, sex, and income level, to make *realized access* measures more nuanced.

Another concern is the modifiable areal unit problem (MAUP) which occurs when using different units to analyze data, potentially generating different results (Openshaw 1984). It is important to note that our results may differ if we were to conduct this analysis using buffer zones and block groups instead of hexagons. One issue with using network data and traditional administrative units, such as census tracts or block groups, is that these units are often defined with major roads along their boundaries, which can complicate transportation analysis. For example, if a trip occurs on a network segment that lies between two block groups, it becomes challenging to assign the trip to a specific block group. Hexagons are commonly used to address this problem because they form a regular, contiguous grid, eliminating the issue of major roads being on the boundaries of the units (Puu 2005).

Equity is supposed to be a prioritized goal of the city master plan (Campbell 1996), which defines equity as providing equal access to active transportation facilities (Lee et al. 2017). As accessibility has been a hotspot for research in GIScience over the past two decades, most studies have focused on the development of tools or models for measuring accessibility (Geertman and Ritsema Van Eck 1995, Miller and Wu 2000, Yigitcanlar et al. 2007). We propose *realized access* to measure the difference within cities in Santa Barbara County and explore the association between *realized access* and two socio-economic factors – median household income and the percentage of the Hispanic population, to demonstrate the importance of considering *realized access* in equity studies. Beyond these two factors, integrating additional socioeconomic data, such as gender, family composition, workplace location and type, could help uncover disparities in accessibility. For example, Hispanic populations residing in Lompoc and Santa Maria might need to commute longer distances, such as to Santa Barbara, for work. Therefore, it is not actually feasible for them to commute by bike. Even with access to biking infrastructure, they might not utilize it for commuting purposes. Future work could examine other biking behaviors beyond commuting, investigate emerging mobility data across broader regions, and incorporate more social and economic aspects of access to advance equity studies.

## 5. Conclusions

*Realized* accessibility in GIScience includes determinants and constraints from natural, built, and perceptual environments. The application of *realized access* measures allows new mobility data to reflect social and economic aspects of access, which are often underestimated in *idealized access* measures. We conduct a case study focusing on bicycling access in Santa Barbara County, calculating both *idealized and realized access*. We find differences in accessibility between *idealized* and *realized access* in some regions. Additionally, cities with increased bicycling access in *realized* measures have higher median household incomes and lower Hispanic population percentages, and vice versa. As novel mobility data become more widely available, advancing proximity measures of accessibility is essential, particularly for equity analysis. We encourage the application of *realized access* measures to evaluate accessibility in other regions and further exploration of new measures of accessibility with emerging mobility data.

Cracu, G.-M., Schvab, A., Prefac, Z., Popescu, M., & Sîrodoev, I. (2024). A GIS-based assessment of pedestrian accessibility to urban parks in the city of Constanța, Romania. *Applied Geography*, *165*, 103229. https://doi.org/10.1016/j.apgeog.2024.103229

Delamater, P. L., Messina, J. P., Shortridge, A. M., & Grady, S. C. (2012). Measuring geographic access to health care: Raster and network-based methods. *International Journal of Health Geographics*, *11*(1), 15. https://doi.org/10.1186/1476-072X-11-15

Dodge, S., & Nelson, T. A. (2023). A framework for modern time geography: Emphasizing diverse constraints on accessibility. *Journal of Geographical Systems*, *25*(3), 357–375. https://doi.org/10.1007/s10109-023-00404-1

Eckert, J., & Shetty, S. (2011). Food systems, planning and quantifying access: Using GIS to plan for food retail. *Applied Geography*, *31*(4), 1216–1223. https://doi.org/10.1016/j.apgeog.2011.01.011

Eldeeb, G., Mohamed, M., & Páez, A. (2021). Built for active travel? Investigating the contextual effects of the built environment on transportation mode choice. *Journal of Transport Geography*, *96*, 103158. https://doi.org/10.1016/j.jtrangeo.2021.103158

El-Geneidy, A., Grimsrud, M., Wasfi, R., Tétreault, P., & Surprenant-Legault, J. (2014). New evidence on walking distances to transit stops: Identifying redundancies and gaps using variable service areas. *Transportation*, *41*(1), 193–210. https://doi.org/10.1007/s11116-013-9508-z

Evans, L. (2003). Transportation Safety. In R. W. Hall (Ed.), *Handbook of Transportation Science* (pp. 67–112). Springer US. https://doi.org/10.1007/0-306-48058-1_4

Ewing, R., & Handy, S. (2009). Measuring the Unmeasurable: Urban Design Qualities Related to Walkability. *Journal of Urban Design*, *14*(1), 65–84. https://doi.org/10.1080/13574800802451155

Ferster, C., Nelson, T., Laberee, K., & Winters, M. (2021). Mapping bicycling exposure and safety risk using Strava Metro. *Applied Geography*, *127*, 102388. https://doi.org/10.1016/j.apgeog.2021.102388
22

Firth, C. L., Hosford, K., & Winters, M. (2021). Who were these bike lanes built for? Social-spatial inequities in Vancouver's bikeways, 2001–2016. *Journal of Transport Geography*, *94*, 103122. https://doi.org/10.1016/j.jtrangeo.2021.103122

Fischer, J., Nelson, T., & Winters, M. (2020). Comparing Spatial Associations of Commuting versus Recreational Ridership Captured by the Strava Fitness App. *Findings*. https://doi.org/10.32866/001c.16710

Frater, J., & Kingham, S. (2018). Gender equity in health and the influence of intrapersonal factors on adolescent girls' decisions to bicycle to school. *Journal of Transport Geography*, *71*, 130–138. https://doi.org/10.1016/j.jtrangeo.2018.07.011

Furth, P. G., Sadeghinasr, B., & Miranda-Moreno, L. (2023). Slope stress criteria as a complement to traffic stress criteria, and impact on high comfort bicycle accessibility. *Journal of Transport Geography*, *112*, 103708. https://doi.org/10.1016/j.jtrangeo.2023.103708

Geertman, S. C. M., & Ritsema Van Eck, J. R. (1995). GIS and models of accessibility potential: An application in planning. *International Journal of Geographical Information Systems*, *9*(1), 67–80. https://doi.org/10.1080/02693799508902025

Gilderbloom, J., Grooms, W., Mog, J., & Meares, W. (2016). The green dividend of urban biking? Evidence of improved community and sustainable development. *Local Environment*, *21*(8), 991–1008. https://doi.org/10.1080/13549839.2015.1060409

Gillies, Andrew. (2022). "State Commission Funds a Historic $80 Million for Santa Barbara Transportation Projects." *News Channel 3-12*.

Grudgings, N., Hughes, S., & Hagen-Zanker, A. (2021). The comparison and interaction of age and gender effects on cycling mode-share: An analysis of commuting in England and Wales. *Journal of Transport & Health*, *20*, 101004. https://doi.org/10.1016/j.jth.2020.101004

Gupta, M. (2024). *Power To Pedal: A Gendered Analysis of the Barriers and Joys of Cycling in Oakland*. https://doi.org/10.7922/G2FF3QQ2
23

Mueller, A. G., & Trujillo, D. J. (2019). The Impact of Zoning and Built Environment Characteristics on Transit, Biking, and Walking. *Journal of Sustainable Real Estate*, *11*(1), 108–129. https://doi.org/10.22300/1949-8276.11.1.108

Mueller, N., Rojas-Rueda, D., Cole-Hunter, T., de Nazelle, A., Dons, E., Gerike, R., Götschi, T., Int Panis, L., Kahlmeier, S., & Nieuwenhuijsen, M. (2015). Health impact assessment of active transportation: A systematic review. *Preventive Medicine*, *76*, 103–114. https://doi.org/10.1016/j.ypmed.2015.04.010

Müller, C., Paulsen, L., Bucksch, J., & Wallmann-Sperlich, B. (2024). Built and natural environment correlates of physical activity of adults living in rural areas: A systematic review. *International Journal of Behavioral Nutrition and Physical Activity*, *21*(1), 52. https://doi.org/10.1186/s12966-024-01598-3

O'Neill, W. A., Ramsey, R. D., & Chou, J. (1992). Analysis of transit service areas using geographic information systems. *Transportation Research Record*, *364*, 131.

Openshaw, S. (1984). The Modifiable Areal Unit Problem. *Concepts and Techniques in Modern Geography*. https://cir.nii.ac.jp/crid/1570291225725496704

Pearce, J., Witten, K., & Bartie, P. (2006). Neighbourhoods and health: A GIS approach to measuring community resource accessibility. *Journal of Epidemiology & Community Health*, *60*(5), 389–395. https://doi.org/10.1136/jech.2005.043281

Piatkowski, D. P., & Marshall, W. E. (2015). Not all prospective bicyclists are created equal: The role of attitudes, socio-demographics, and the built environment in bicycle commuting. *Travel Behaviour and Society*, *2*(3), 166–173. https://doi.org/10.1016/j.tbs.2015.02.001

Pot, F. J., van Wee, B., & Tillema, T. (2021). Perceived accessibility: What it is and why it differs from calculated accessibility measures based on spatial data. *Journal of Transport Geography*, *94*, 103090. https://doi.org/10.1016/j.jtrangeo.2021.103090

Puu, T. (2005). On the Genesis of Hexagonal Shapes. *Networks and Spatial Economics*, *5*(1), 5–20. https://doi.org/10.1007/s11067-005-6659-2

Reynolds, C. C., Harris, M. A., Teschke, K., Cripton, P. A., & Winters, M. (2009). The impact of transportation infrastructure on bicycling injuries and crashes: A review of the literature. *Environmental Health*, *8*(1), 47. https://doi.org/10.1186/1476-069X-8-47

Rodriguez-Valencia, A., Rosas-Satizábal, D., Gordo, D., & Ochoa, A. (2019). Impact of household proximity to the cycling network on bicycle ridership: The case of Bogotá. *Journal of Transport Geography*, *79*, 102480. https://doi.org/10.1016/j.jtrangeo.2019.102480

Ross, A., Rodríguez, A., & Searle, M. (2017). Associations between the Physical, Sociocultural, and Safety Environments and Active Transportation to School. *American Journal of Health Education*, *48*(3), 198–209. https://doi.org/10.1080/19325037.2017.1292877

Schipperijn, J., Bentsen, P., Troelsen, J., Toftager, M., & Stigsdotter, U. K. (2013). Associations between physical activity and characteristics of urban green space. *Urban Forestry & Urban Greening*, *12*(1), 109–116. https://doi.org/10.1016/j.ufug.2012.12.002

*SciELO - Brazil—Por onde as mulheres escolhem caminhar? Segurança feminina em espaços públicos Por onde as mulheres escolhem caminhar? Segurança feminina em espaços públicos*. (n.d.). Retrieved June 11, 2024, from https://www.scielo.br/j/cm/a/b3nRJYZhb8CsfT8QvvRtZkt/?lang=en

Sener, I. N., Lee, K., Hudson, J. G., Martin, M., & Dai, B. (2021). The challenge of safe and active transportation: Macrolevel examination of pedestrian and bicycle crashes in the Austin District. *Journal of Transportation Safety & Security*, *13*(5), 525–551. https://doi.org/10.1080/19439962.2019.1645778

Stefansdottir, H., Næss, P., & Ihlebæk, C. M. (2019). Built environment, non-motorized travel and overall physical activity. *Travel Behaviour and Society*, *16*, 201–213. https://doi.org/10.1016/j.tbs.2018.08.004

Steinmann, J., Wilson, B., McSweeney, M., Bandoles, E., & Hayhurst, L. M. C. (2021). *An Exploration of Safe Space: From a Youth Bicycle Program to the Road*. https://doi.org/10.1123/ssj.2020-0155

Stewart, O. (2011). Findings from Research on Active Transportation to School and Implications for Safe Routes to School Programs. *Journal of Planning Literature*, *26*(2), 127–150. https://doi.org/10.1177/0885412210385911
28

**Acknowledges**

We gratefully acknowledge Strava for their support in collecting and compiling origins and destinations data. We also extend our gratitude to the users who contributed their data for this research.